\documentclass[journal]{IEEEtran}

\newcommand{\ie}{{\em i.e., }}
\newcommand{\eg}{{\em e.g., }}

\usepackage{cite}
\usepackage{graphicx,color,subfigure}
\usepackage{graphics,url}
\usepackage[draft]{hyperref}
\usepackage{caption}
\usepackage{amsmath}
\usepackage{algorithm}
\usepackage[noend]{algpseudocode}
\usepackage[nolist]{acronym}
\usepackage{fancyvrb}
\usepackage{listings}
\usepackage{adjustbox}
\usepackage{tabularx} 
\PassOptionsToPackage{table,xcdraw}{xcolor}
\usepackage{color, colortbl}
\usepackage{graphicx}

\newcolumntype{L}[1]{>{\raggedright\let\newline\\\arraybackslash\hspace{0pt}}m{#1}}
\newcolumntype{C}[1]{>{\centering\let\newline\\\arraybackslash\hspace{0pt}}m{#1}}
\newcolumntype{R}[1]{>{\raggedleft\let\newline\\\arraybackslash\hspace{0pt}}m{#1}}

\usepackage{amssymb}
\usepackage{pifont}
\usepackage{etoolbox}\AtBeginEnvironment{algorithmic}{\small}
\algnewcommand\algorithmicforeach{\textbf{for each}}
\algdef{S}[FOR]{ForEach}[1]{\algorithmicforeach\ #1\ \algorithmicdo}



\makeatother
\algnewcommand{\LineComment}[1]{\State \(\triangleright\) #1}

\usepackage{amsthm}
\newtheorem{thm}{Theorem}

\newtheorem{lem}[thm]{Lemma}

\newtheorem{defn}[thm]{Definition}

\def\equationautorefname~#1\null{%
  (#1)\null
}

\begin{document}
%
\title{Clear as MUD: Generating, Validating \\and Applying IoT Behaviorial Profiles \\ (Technical Report)}


\author{\IEEEauthorblockN{Ayyoob Hamza\IEEEauthorrefmark{1},
Dinesha Ranathunga\IEEEauthorrefmark{2},
H. Habibi Gharakheili\IEEEauthorrefmark{1} 
Matthew Roughan\IEEEauthorrefmark{1} and
Vijay Sivaraman\IEEEauthorrefmark{1}}
\IEEEauthorblockA{\IEEEauthorrefmark{1}University of New South Wales, Australia}
\IEEEauthorblockA{\IEEEauthorrefmark{2}ACEMS, University of Adelaide, Australia\\
Email: ayyoobhamza@student.unsw.edu.au, \{h.habibi, vijay\}@unsw.edu.au,\\ \{dinesha.ranathunga, matthew.roughan\}@adelaide.edu.au}}

\maketitle

\thispagestyle{plain}
\pagestyle{plain}

\begin{abstract}
IoT devices are increasingly being implicated in cyber-attacks, driving community concern about the risks they pose to critical infrastructure, corporations, and citizens. In order to reduce this risk, the IETF is pushing IoT vendors to develop formal specifications of the intended purpose of their IoT devices, in the form of a Manufacturer Usage Description (MUD), 
so that their network behavior in any operating environment can be locked down and verified rigorously.

This paper aims to assist IoT manufacturers in developing and verifying MUD profiles, while also helping adopters of these devices to ensure they are compatible with their organizational policies. Our first contribution is to develop a tool that takes the traffic trace of an arbitrary IoT device as input and automatically generates a MUD profile for it. We contribute our tool as open source, apply it to 28 consumer IoT devices, and highlight insights and challenges encountered in the process. Our second contribution is to apply a formal semantic framework that not only validates a given MUD profile for consistency, but also checks its compatibility with a given organizational policy. Finally, we apply our framework to representative organizations and selected devices, to demonstrate how MUD can reduce the effort needed for IoT acceptance testing. 
\end{abstract}



\IEEEpeerreviewmaketitle

\section{Introduction}\label{sec:intro}
Many online IoT devices can be found on search engines such as Shodan \cite{Shodan}, and if they have any vulnerability, it can then be exploited at scale, for example, to launch DDoS attacks. For instance, Dyn, a major DNS, was attacked by means of a DDoS attack originated from a large IoT botnet composed of thousands of compromised IP-cameras \cite{Dyn16}.
IoT devices, exposing TCP/UDP ports to arbitrary remote endpoints, are used by attackers to reflect/amplify attacks
or to infiltrate otherwise secure networks.

This has prompted standards bodies to provide guidelines for the Internet community to build secure IoT devices and services \cite{DHS16,NIST16,ENISA17}. In addition, the US Federal Communications Commission (FCC) has stated the need for additional regulation of IoT systems \cite{FCC16}. There is also a practical proposal called Manufacturer Usage Description (MUD) which is being reviewed by the IETF. This proposal requires manufacturers of IoTs to publish a behavioral profile of their device. Manufacturers have the most insight into what resources the devices will need once they are installed in a network; for example, an IP camera should only use DNS and DHCP on the local network, and communicate to NTP server and its own cloud-based controller, but nothing else. 
These requirements vary across IoTs from different manufacturers. Knowing each one's requirements would allow a tight set of \ac{ACLs} to be imposed.

Manufacturers should be best-suited to author the network profiles which will be required by their devices. Therefore, the IETF MUD proposal provides a light-weight model of achieving very effective baseline security for IoT devices by simply allowing a network to automatically configure the required network access for IoT devices so that they can perform their intended functions without granting them unrestricted network privileges.
However, what do we do if a manufacturer incorrectly specifies its MUD or the MUD profile conflicts with the security policy of our network? 

The MUD approach differs from existing cyber-security approaches because it is an open and standards-based approach, uses the subject matter expertise of device manufacturers, and importantly is scalable to minimize the effort of securing IoT endpoints. 
But, manufacturers need help to generate and verify their MUD profiles, and network operators to verify the compatibility of MUD profiles with their organizational policies.  

MUD is a new and emerging paradigm, and there is little collective wisdom today on how manufacturers should develop behavioral profiles of their IoT devices, or how organizations should use these profiles to secure their network. This paper is our attempt to address both these shortcomings.

Our specific contributions are as follows: We first develop a tool that takes the packet trace of an IoT device as input and automatically generates a MUD profile for it. We contribute our tool as open source, apply it to 28 consumer IoT devices, and highlight insights and challenges encountered in the process. We then apply a formal semantic framework that not only validates a given MUD profile for consistency, but also checks its compatibility with a given organizational policy. Finally, we apply our framework to representative organizations and selected devices, to demonstrate how MUD can reduce the effort needed for IoT acceptance testing.

\section{Background and Related Works}\label{sec:prior}

Security of IoT devices has not kept up with the rapid pace of innovation, creating substantial safety and economic risks for the Internet \cite{Survey17}. 
Today, many IoT products do not incorporate even basic security measures \cite{IoTSnp17}. 
As IoT botnets grow and become mature, attackers are using them to launch more advanced DDoS attacks \cite{f5Labs17}; devices such as baby monitors, refrigerators and smart plugs have been hacked and controlled remotely \cite{sivaraman2016smart}; there are many cameras that can be accessed publicly \cite{SonyCam,Insecam18} since network operators often do not impose access control policies to these IoT devices \cite{CiscoReport18}. 

While it may seem impossible to identify all potential forms of adversary or to stop all anticipated threats, we should at least be able to enhance the penetration resistance of IoTs and make them less vulnerable \cite{NIST16}.
It has been shown that IoTs often employ a limited set of flow rules to few remote/Internet-based endpoints in a recognizable pattern \cite{SmartCity17}.
So it should be possible to discover and prevent unintended activity at the network level. But, only if we knew what behavior is intended.

\begin{figure}[t]
\centering
\includegraphics[scale=0.2]{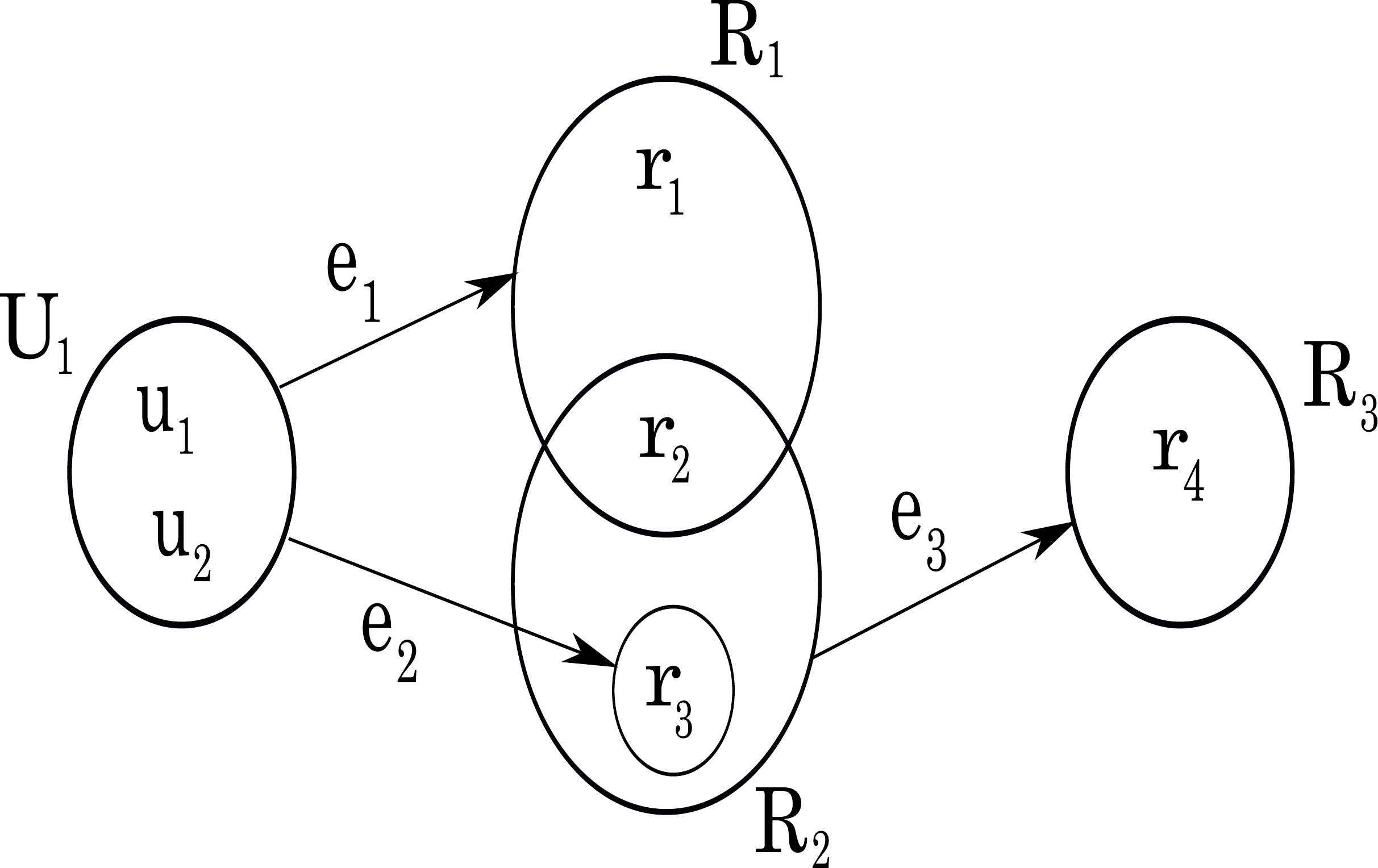}
\caption{A metagraph consisting of six variables, five sets and three edges.}
\label{fig:mg1}
\end{figure}

The Internet draft for MUD specification \cite{ietfMUD18} promotes the device manufacturer to clearly define the communication pattern of a device in the form of access control rules. This allows network monitoring tools to secure the device based on the principle that if an observed behavior of the device is not in the profile, then that communication is considered a threat. By translating access control rules to network rules we can restrict local/remote communication to restricted flows and endpoints. This helps safeguard the device.

An operational critical infrastructure or enterprise network, should have a local security policy. The IETF MUD specification proposes plug and play of IoT devices in a network, but this convenience brings with it serious security implications. For instance, a MUD controller's attempt to dynamically enable a MUD policy in a network for a freshly connected IoT device can result in the breach of security policy or best practices. In this paper we propose an extension that can be used along with the MUD controller to check compliance of a MUD profile with
an organizational policy. We show how such checks can be automated and constructed using rigorous formal semantics, before devices are powered or deployed.

Lack of formal policy modeling capabilities in current network configuration systems 
contribute to the frequent misconfigurations found  \cite{wool2010,ranathunga2016T,ranathunga2016G}.

A {\em metagraph} is a generalized graph theoretic structure that offers rigorous formal foundations
for modeling and analyzing communication-network policies in general. We use them here to model and analyze MUD policies.
A metagraph is a directed graph between a collection of sets of `atomic' elements \cite{basu2007}. 
Each set is a node in the graph and each directed edge represents the relationship between the sets.
\autoref{fig:mg1} shows an example where a set of users ($U_1$) are related to sets of network resources ($R_1$, $R_2$, $R_3$)
by the edges $e_1, e_2$ and $e_3$ describing which user $u_i$ is allowed to access resource $r_j$.

Metagraphs can also have attributes associated with their edges.
An example is a {\em conditional metagraph}
which includes propositions -- statements that may be true or false -- assigned to their edges as qualitative attributes \cite{basu2007}.
The generating sets of these metagraphs are partitioned into a variable set and a proposition set.
A conditional metagraph is formally defined as follows:

\begin{defn}[Conditional Metagraph] 
 A conditional metagraph is a metagraph $S$=$\langle X_p \cup X_v, E \rangle$
 in which $X_p$ is a set of propositions and $X_v$ is a set of variables,
and:\\
\indent 1. at least one vertex is not null, \ie $\forall e' \in E, V_{e'} \cup W_{e'} \neq \phi$\\ 
\indent 2. the invertex and outvertex of each edge must be disjoint, \ie $X = X_v \cup X_p$ with $X_v \cap X_p = \phi$\\
\indent 3. an outvertex containing propositions cannot contain other elements, \ie 
$\forall p \in X_p, \forall e' \in E$, if $p \in W_{e'}$, then $W_{e'}={p}$. 
 \label{def:metagraph}
\end{defn}

Conditional metagraphs enable the specification of stateful network-policies
and have several useful operators. 
These operators readily allow one to analyze MUD policy properties like consistency.
To the best of our knowledge, this is the first attempt for automatic generation of MUD profile and also formally check for consistency
and its compatibility with an organizational policy prior to installing the access control rules.

\section{MUD Profile Generation}\label{sec:mud}

\begin{table}[t!]
	\small
	\caption{Flows observed for Blipcare BP monitor (*: wildcard, proto: Protocol, sPort: source port number, dPort: destination port number).} 
	\vspace{-2mm}
	\centering 
	\begin{adjustbox}{max width=0.48\textwidth}
		\begin{tabular}{l l l l l} 
			\hline\hline 
			\multicolumn{1}{p{0.7cm}}{\raggedleft \textbf{Source} } & \multicolumn{1}{p{0.7cm}}{\raggedleft \textbf{Destination}} & 
			\multicolumn{1}{p{0.75cm}}{\raggedleft \textbf{proto}} & \multicolumn{1}{p{0.85cm}}{\raggedleft \textbf{sPort} } & 
			\multicolumn{1}{p{0.85cm}}{\raggedleft \textbf{dPort}}\\
			\hline 
			* & 192.168.1.1 & 17 & * & 53\\
			192.168.1.1 & * & 17 & 53 & * \\
			* & tech.carematix.com & 6 & * & 8777\\
			tech.carematix.com & * & 6 & 8777 & *\\
			\hline 
		\end{tabular}
	\end{adjustbox}
	\label{table:bpmrules} 
	\vspace{-5mm}
\end{table}

The IETF MUD specification is still evolving as a draft. Hence, IoT device manufacturers have not yet provided MUD profiles for their devices. 
We, therefore, developed a tool -- \textit{MUDgee} -- which automatically generates a MUD profile for an IoT device
from its traffic trace in order to make this process faster, cheaper and more accurate.
In this section, we describe the structure of our open source tool \cite{mudgenerator}, 
apply it to traces of 28 consumer IoT devices, and highlight insights. 

We captured traffic flows for each IoT device during a six month observation period, to generate our MUD rules.
The rules reflect an application whitelisting model (\ie there are no explicit `drop' rules).
Having a combination of `accept' and `drop' rules requires a notion of rule priority (\ie order) and is not supported by the
current IETF MUD draft.
For example, \autoref{table:bpmrules} shows traffic flows observed in our lab for a Blipcare blood pressure monitor. 
The device only generates traffic whenever it is used. 
It first resolves its intended server at \verb|tech.carematrix.com| by exchanging a DNS query/response with the default gateway (\ie the top two flows).
It then uploads the measurement to its server operating on TCP port 8777 (described by the bottom two rules).

\begin{figure*}[t!]
   \centering
   \includegraphics[scale=0.45]{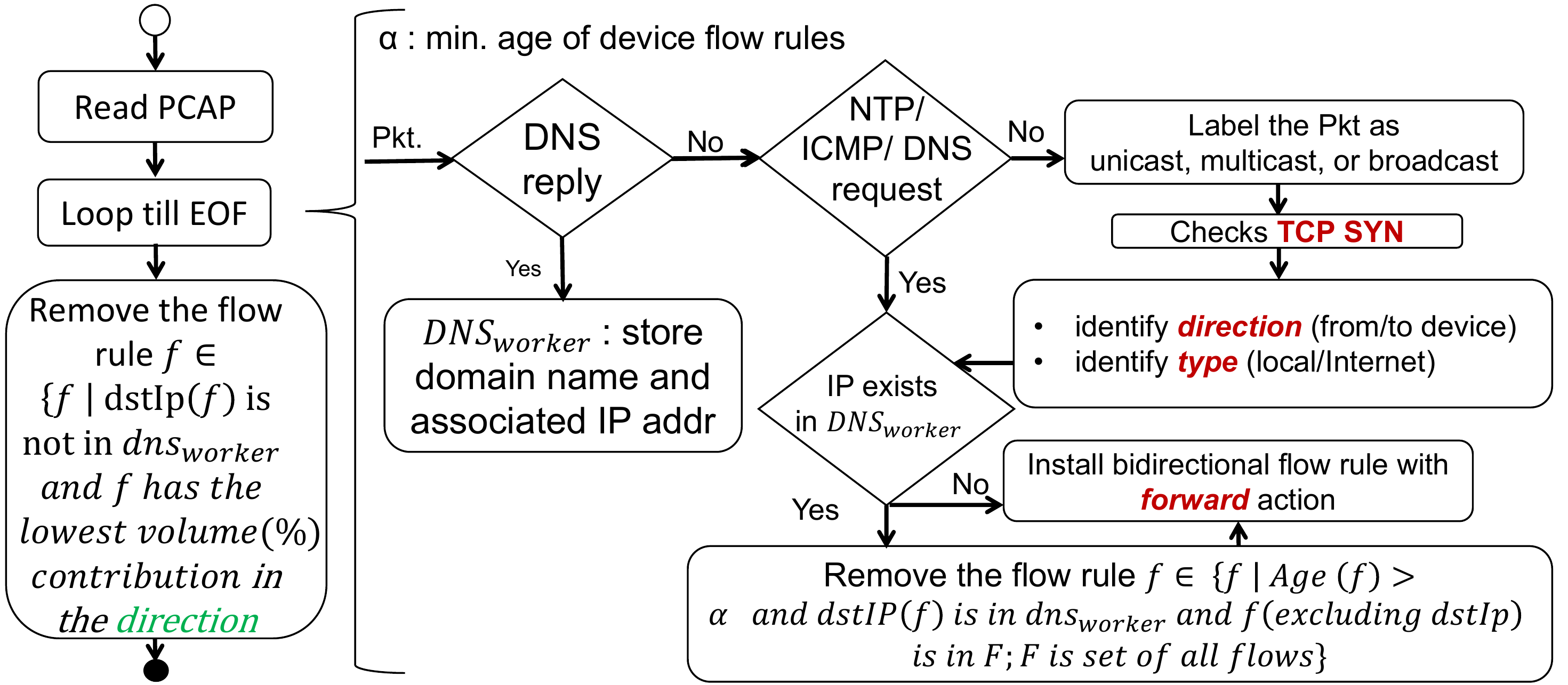}
   \caption{Algorithm for capturing device flows and inserting reactive rules.}
   \label{fig:flowidentifier}
\end{figure*}

\begin{table*}[t]
	\centering
	\caption{Proactive flow rules for an IoT device.}\label{table:devrules}
	\vspace{-0.2cm}
	\begin{adjustbox}{max width=0.80\textwidth}	
	\begin{tabular}{|l|c|c|c|c|c|c|c|c|c|c|}
		\hline 
		\textbf{id} & \textbf{sEth} & \textbf{dEth} & \textbf{typeEth} & \textbf{Source} & \textbf{Destination} & \textbf{proto} & \textbf{sPort} & \textbf{dPort} & \textbf{priority} & \textbf{action}\tabularnewline
		\hline 
		\hline 
		a & $<$\verb|gwMAC|$>$ & $<$\verb|devMAC|$>$  & \verb|0x0800| & {*} & {*} & 1 & {*} & {*} & 100 & forward\tabularnewline
		\hline 
		b.1 & $<$\verb|devMAC|$>$  & $<$\verb|gwMAC|$>$ & \verb|0x0800| & {*} & {*} & 1 & {*} & {*} & 100 & mirror\tabularnewline
		\hline 
		b.2 & $<$\verb|devMAC|$>$  & $<$\verb|gwMAC|$>$ & \verb|0x86dd| & {*} & {*} & 58 & {*} & {*} & 100 & mirror\tabularnewline
		\hline 
		c & $<$\verb|gwMAC|$>$ & $<$\verb|devMAC|$>$  & \verb|0x0800| & {*} & {*} & 17 & 123 & {*} & 100 & forward\tabularnewline
		\hline 
		d.1 & $<$\verb|gwMAC|$>$ & $<$\verb|devMAC|$>$  & {*} & {*} & {*} & 17 & 53 & {*} & 100 & mirror\tabularnewline
		\hline
		d.2 & $<$\verb|devMAC|$>$  & $<$\verb|gwMAC|$>$ & {*} & {*} & {*} & 17 & {*} & 53 & 100 & mirror\tabularnewline
		\hline  		
		
		e.1 & {*} & $<$\verb|devMAC|$>$  & \verb|0x0806| & {*} & {*} & {*} & {*} & {*} & 100 & forward\tabularnewline
		\hline 
		e.2 & $<$\verb|devMAC|$>$  & {*} &  \verb|0x0806| & {*} & {*} & {*} & {*} & {*} & 100 & forward\tabularnewline
		\hline
		f & $<$\verb|gwMAC|$>$ & $<$\verb|devMAC|$>$  & {*} & gw local IP & {*} & * & {*} & {*} & 90 & forward\tabularnewline
		\hline 
		g & $<$\verb|devMAC|$>$  & $<$\verb|gwMAC|$>$ & {*} & {*} & gw local IP & * & {*} & {*} & 90 & forward\tabularnewline
		\hline 
		h & $<$\verb|devMAC|$>$  & {*}  &   \verb|0x888e| & {*} & {*} &{*} & {*} & {*} & 3 & forward \tabularnewline
		\hline
		i & $<$\verb|devMAC|$>$  & {*}  &  {*} & {*} & {*} & {*} & {*} & {*} & 2 & mirror\tabularnewline
		\hline
		j & {*} & $<$\verb|devMAC|$>$ &  {*} & {*} & {*} & {*} & {*} & {*} & 2 & mirror\tabularnewline
		\hline	
	\end{tabular}
	\end{adjustbox}
\end{table*}

\subsection{MUDgee Architecture}

{\em MUDgee} implements a programmable virtual switch (vSwitch) with a header inspection engine attached and plays an input PCAP trace (of an arbitrary IoT device) into the switch. {\em MUDgee} has two separate modules; (a) captures and tracks all TCP/UDP flows to/from device, and (b) composes a MUD profile from the flow rules.   
 
\noindent \textbf{Capture intended flows:} Consumer IoT devices use services provided by remote servers on the cloud and also expose services to local hosts (\eg a mobile App). We track (intended) device activities for both remote and local communications using separate flow rules. 

It is challenging to capture services (\ie especially those operating on non-standard TCP/UDP ports) that a device is either accessing or exposing. This is because local/remote services operate on static port numbers whereas source port numbers are dynamic (and chosen randomly) for different flows of the same service. 
Also, inferring the direction of UDP flows is non trivial, though for TCP flows it can be deduced by inspecting the SYN flag.
We developed an algorithm (Fig.~\ref{fig:flowidentifier}) to capture bidirectional flows for an IoT device. 

We first configure the vSwitch with a set of proactive rules, each with a specific action (\ie``forward'' or ``mirror'') and a priority. 
These rules are listed in \autoref{table:devrules}.
Proactive rules with a `mirror' action will feed the header inspection engine with a copy of the matched packets. 
Our inspection algorithm, shown in Fig.~\ref{fig:flowidentifier}, will insert a corresponding reactive rule into the vSwitch. 

For example, a DNS reply packet is matched to a top priority flow (\ie flow id `d.1' in \autoref{table:devrules})
and our algorithm extracts and stores the domain name and its  associated IP address into a DNS cache table.
This cache is dynamically updated upon arrival of a DNS reply matching an existing request.

The MUD specification requires the segregation of traffic to and from a device for both local and Internet communications.
Our algorithm achieves this by assigning a unique priority to the reactive rules associated with each of the groups: from-local, to-local, from-Internet and to-Internet.
We use a specific priority for flows that contain a TCP SYN to identify if the device or the remote entity initiated the communication.

\autoref{table:awairrules} shows the flow rules captured for an Awair air quality monitor. 
The device uses the local gateway and the Google DNS server (\ie 8.8.8.8) to resolve its domain names. 
It also periodically time synchronizes with \verb|pool.net.org|. 
The IoT device does not have any local communication and only uses TCP for remote communication.

 \begin{table*}[t!]
	\centering
	\caption{Awair air quality reactive flow rules}
	\label{table:awairrules}
	\vspace{-0.2cm}
	\begin{adjustbox}{max width=0.75\textwidth}
	\begin{tabular}{|c|c|c|c|c|c|c|c|c|}
		\hline 
		\textbf{sEth} & \textbf{dEth} & \textbf{typeEth} & \textbf{Source} & \textbf{Destination} & \textbf{proto} & \textbf{sPort} & \textbf{dPort} & \textbf{priority}\tabularnewline
		\hline 
		\hline 
		$<$\verb|devMAC|$>$  & ff:ff:ff:ff:ff:ff & 0x0006 & {*} & {*} & {*} & {*} & {*} & 102\tabularnewline
		\hline 
		$<$\verb|devMAC|$>$  & ff:ff:ff:ff:ff:ff & 0x0800 & {*} & 255.255.255.255 & 17 & {*} & 67 & 102\tabularnewline
		\hline 
		$<$\verb|devMAC|$>$  & $<$\verb|gwMAC|$>$ &  \verb|0x0800| & {*} & 8.8.8.8 & 17 & {*} & 53 & 101\tabularnewline
		\hline 
		$<$\verb|devMAC|$>$  & $<$\verb|gwMAC|$>$ &  \verb|0x0800| & {*} & 192.168.1.1 & 17 & {*} & 53 & 101\tabularnewline
		\hline 
		$<$\verb|devMAC|$>$  & $<$\verb|gwMAC|$>$ &  \verb|0x0800| & {*} & pool.ntp.org & 17 & {*} & 123 & 101\tabularnewline
		\hline 
		$<$\verb|gwMAC|$>$ & $<$\verb|devMAC|$>$  & \verb|0x0800| & 8.8.8.8 & {*} & 17 & 53 & {*} & 101\tabularnewline
		\hline 
		$<$\verb|gwMAC|$>$ & $<$\verb|devMAC|$>$  & \verb|0x0800| & 192.168.1.1 & {*} & 17 & 53 & {*} & 101\tabularnewline
		\hline 
		$<$\verb|gwMAC|$>$ & $<$\verb|devMAC|$>$  & \verb|0x0800| & \verb|pool.ntp.org| & {*} & 17 & 123 & {*} & 101\tabularnewline
		\hline 
		$<$\verb|devMAC|$>$  & $<$\verb|gwMAC|$>$ &  \verb|0x0800| & {*} & ota.awair.is & 6 & {*} & 443 & 32\tabularnewline
		\hline 
		$<$\verb|devMAC|$>$  & $<$\verb|gwMAC|$>$ & \verb|0x0800| & {*} & timeserver.awair.is & 6 & {*} & 443 & 32\tabularnewline
		\hline 
		$<$\verb|devMAC|$>$  & $<$\verb|gwMAC|$>$ & \verb|0x0800| & {*} & api.awair.is & 6 & {*} & 443 & 32\tabularnewline
		\hline 
		$<$\verb|devMAC|$>$  & $<$\verb|gwMAC|$>$ & \verb|0x0800| & {*} & messaging.awair.is & 6 & {*} & 8883 & 32\tabularnewline
		\hline 
		$<$\verb|gwMAC|$>$ & $<$\verb|devMAC|$>$  & \verb|0x0800| & \verb|ota.awair.is| & {*} & 6 & 443 & {*} & 21\tabularnewline
		\hline 
		$<$\verb|gwMAC|$>$ & $<$\verb|devMAC|$>$  & \verb|0x0800| & \verb|timeserver.awair.is| & {*} & 6 & 443 & {*} & 21\tabularnewline
		\hline 
		$<$\verb|gwMAC|$>$ & $<$\verb|devMAC|$>$  & \verb|0x0800| & \verb|api.awair.is| & {*} & 6 & 443 & {*} & 21\tabularnewline
		\hline 
		$<$\verb|gwMAC|$>$ & $<$\verb|devMAC|$>$ & \verb|0x0800| & \verb|messaging.awair.is| & {*} & 6 & 8883 & {*} & 21\tabularnewline
		\hline 
	\end{tabular}
	\end{adjustbox}
\end{table*}

\noindent \textbf{Flow translation to MUD:}
{\em MUDgee} uses the captured traffic flows to generate the MUD profiles for the devices
by considering the following:

\textit{\textbf{Consideration 1}}: We use the DNS cache to reverse lookup the IP address of the remote endpoint to a domain name, if any.

\textit{\textbf{Consideration 2}}: Some consumer IoTs, especially IP cameras, typically use the \ac{STUN} protocol to verify that the user's mobile app can stream video directly from the camera over the Internet. If a device uses the STUN protocol over UDP, we must allow all UDP traffic to/from Internet servers because the STUN servers often require the client device to connect to different IP addresses or port numbers. 

\textit{\textbf{Consideration 3}}: We observed that several smart IP cameras communicate with many remote servers operating on the same port (\eg August doorbell camera). However, no DNS responses were found corresponding to the server IP addresses. So, the device must obtain the IP address of its servers via a non-standard channel (\eg the current server may instruct the device with the IP address of the subsequent server).  For this case, we allow remote traffic to/from any IP addresses (\ie *), but use a specific port number.

\textit{\textbf{Consideration 4}}: Some devices (\eg TPLink plug) use the default gateway as the DNS resolver, and others (\eg Belkin WeMo motion) continuously ping the default gateway. 
The existing MUD draft maps local communication to fixed IP addresses through the controller construct.
We consider the local gateway to act as the controller, and use the name-space \verb|urn:ietf:params:mud:| \verb|gateway| for the gateway. 

\smallskip
\noindent The generated MUD profiles of the 28 consumer IoT devices we analyzed are listed in \autoref{table:IoTdevices} and are publicly available at:
 \verb|http://149.171.189.1/mud/|. 

\subsection{Insights and challenges} 
\begin{figure*}[t]
  \captionsetup{aboveskip=8pt}
  \subfigure[Blipcare BP monitor.]
  {
	{\includegraphics[width=0.47\textwidth,height=0.2\textheight]{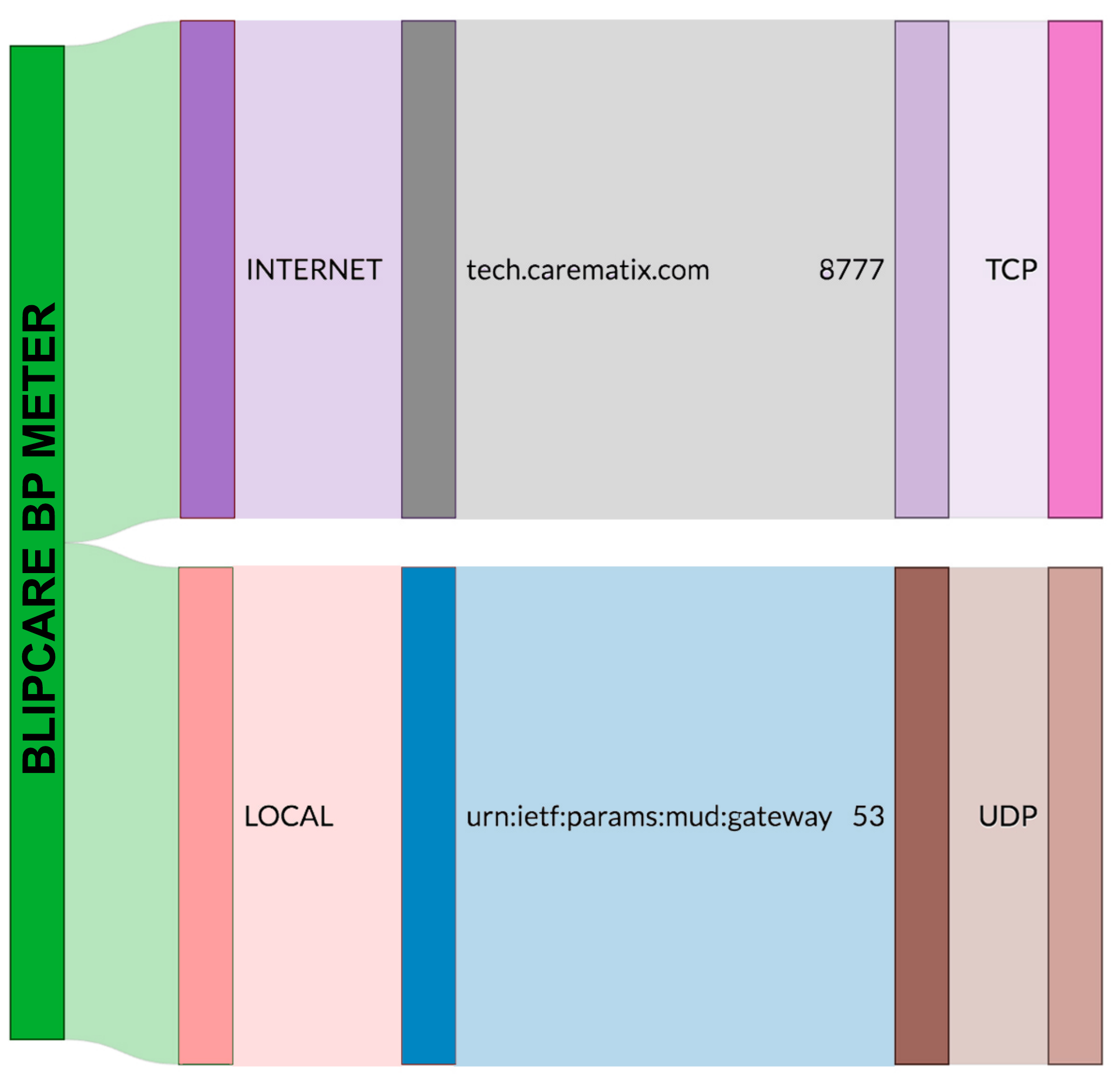}}
	\label{fig:sankeyBP}
 }
  \subfigure[TP-Link camera.]
  {
    {\includegraphics[width=0.47\textwidth,height=0.2\textheight]{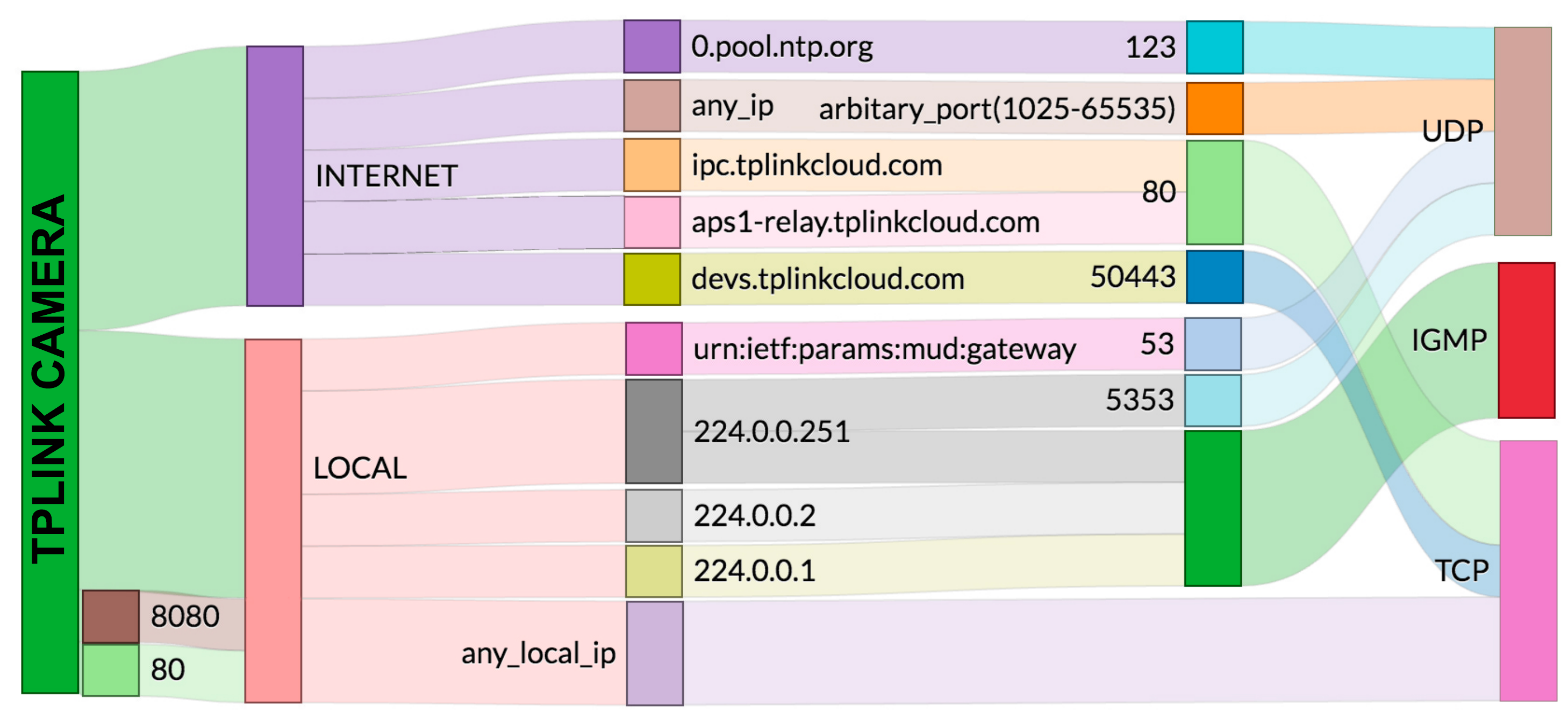}} 
    \label{fig:sankeyTPlinkCam}
 }
  \centering
  \subfigure[Amazon Echo.]
  {
    {\includegraphics[width=0.47\textwidth,height=0.2\textheight]{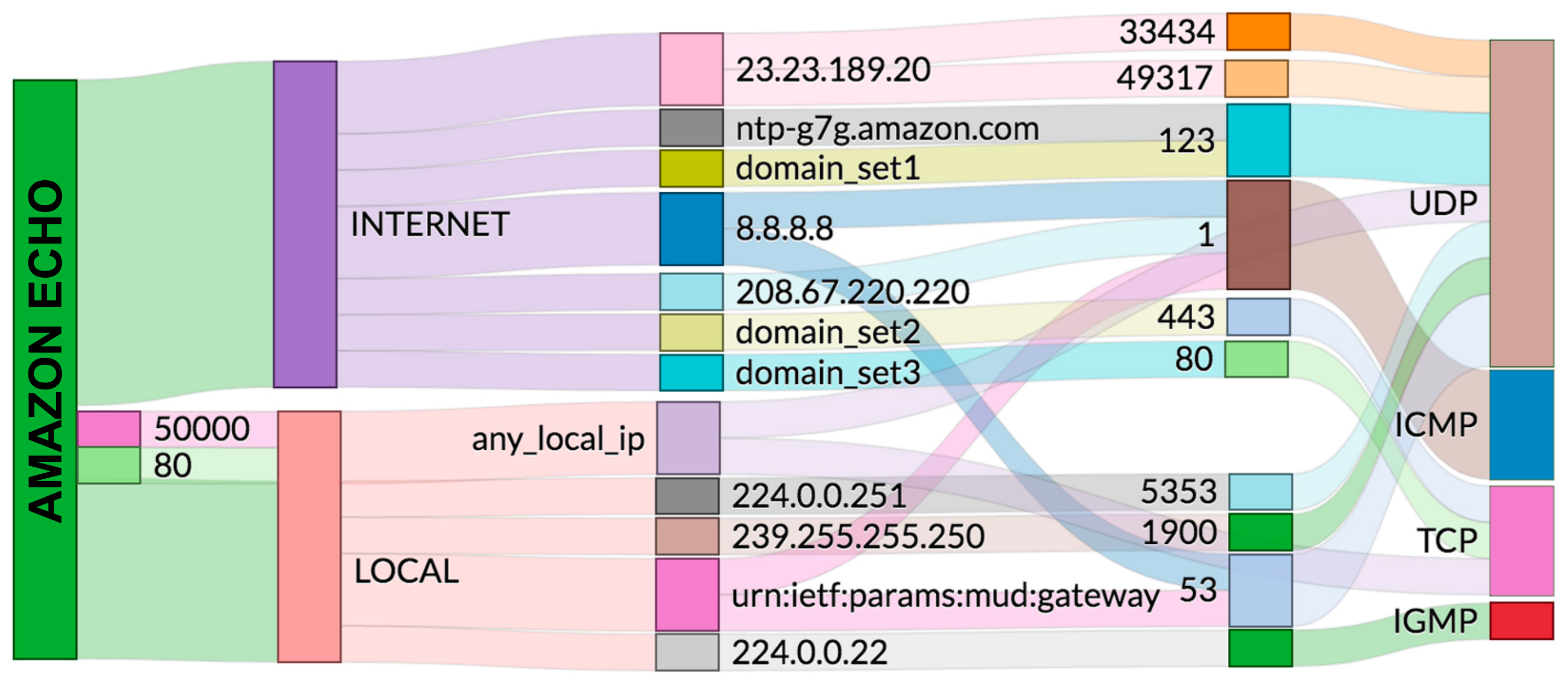}} 
    \label{fig:sankeyEcho}
  }			
 \caption{Sankey diagram of MUD profile for: (a) Blipcare BP monitor, (b)  TP-Link camera, and (c) Amazon Echo.}
  \label{fig:Snakey}
\end{figure*}

The Sankey diagrams \cite{schmidt2008} in Figure~\ref{fig:Snakey} represent the MUD profiles for three IoT devices,
namely Blipcare BP monitor, TP-Link camera and Amazon Echo.

The Blipcare BP monitor is an example device with static functionalities. 
It exchanges DNS queries/responses with the local gateway 
and communicates with a single domain name over TCP port 8777. 
So its behavior can be locked down to a limited set of static flow rules. 
The majority of IoT devices that we tested (\ie 22 out of 28) fall into this category (marked in green in \autoref{table:IoTdevices}).      

Fig.~\ref{fig:sankeyTPlinkCam} exemplifies the second category of our generated MUD profiles.
The TP-Link camera accesses/exposes limited ports on the local network.
It gets its DNS queries resolved, discovers local network using mDNS service over UDP 5353, probes members of certain multicast groups using IGMP, and exposes two TCP ports 80 (management console) and 8080 (unicast video streaming) to local devices. 
All these activities can be defined by a tight set of ACLs. 
But, over the Internet, the camera communicates to its STUN server (accessing an arbitrary range of IP addresses and port numbers shown by the top flow), to port numbers on specific endpoints including time synchronization with \verb|pool.ntp.org|. 
Such IoT devices with static functionalities that are loosely defined, due to use of STUN protocol fall in to this second category (marked in blue in \autoref{table:IoTdevices}).
This category device manufacturers can configure their STUN servers to use a specific set of endpoints and port numbers, instead of a wide and arbitrary range.    

Amazon Echo, represents devices with complex and dynamic functionalities augmentable using custom recipes or skills. 
Such devices (marked in red in \autoref{table:IoTdevices}), 
are able to communicate with a growing range of endpoints on the Internet, 
which the original manufacturer cannot define in advance. 
For example, our Amazon Echo communicates with \verb|meethue.com| over TCP 443, to interact with the Hue lightbulb in the test bed. 
It can also contact the news website \verb|abc.net.au| when prompted by the user. 
For these type of devices, the biggest challenge is how manufacturers can dynamically update their MUD profiles to match the device capabilities.

\begin{table}[t]
	\small
	\caption{List of IoT devices for which we have generated MUD profiles. Devices with purely static functionality are marked in green. Devices with static functionality that is loosely defined (\eg due to use of STUN protocol) are marked in blue. Devices with complex and dynamic functionality are marked in red.} 
	\vspace{-2mm}
	\centering 
	\begin{adjustbox}{max width=0.495\textwidth}
		\begin{tabular}{|L{2cm}|L{6cm}|}
			\hline 
			\multicolumn{1}{|c|} {\textbf{Type}} & 	\multicolumn{1}{|c|} {\textbf{IoT device}}\tabularnewline
			\hline 
			Camera & {\color{green}Netatmo Welcome, Dropcam, Withings Smart Baby Monitor, Canary camera}, {\color{blue}TP-Link Day Night Cloud camera, August doorbell camera, Samsung SmartCam, Ring doorbell, Belkin NetCam}  \tabularnewline
			\hline 
			Air quality sensors & {\color{green}Awair air quality monitor, Nest smoke sensor, Netatmo weather station}  \tabularnewline
			\hline 	
			Healthcare devices & {\color{green}Withings Smart scale, Blipcare Blood Pressure meter, Withings Aura smart sleep sensor}  \tabularnewline
			\hline 			
			Switches and Triggers & {\color{green}iHome power plug, WeMo power switch, TPLink plug, Wemo Motion Sensor} \tabularnewline
			\hline 		
			Lightbulbs & {\color{green}Philips Hue lightbulb, LiFX bulb} \tabularnewline
			\hline 				
			Hub & {\color{red}Amazon Echo}, {\color{green}SmartThings}\tabularnewline
			\hline
			Multimedia& {\color{green}Chromecast}, {\color{red}Triby Speaker}\tabularnewline
			\hline
			Other & {\color{green}HP printer, Pixstar Photoframe, Hello Barbie} \tabularnewline
			\hline 		
		\end{tabular}
	\end{adjustbox}
        \vspace{-6mm}
	\label{table:IoTdevices} 
\end{table}

\section{MUD profile verification}
\label{sec:pdn}

Network operators should not automatically implement a 
device's MUD policy without first checking its validity.
Lack of rigorous policy verification in current network-configuration systems
is a key contributor to the configuration errors commonly found \cite{wool2010,ranathunga2016T}.
We describe here in detail, the verification steps
we apply on a device MUD profile to ensure only policies that
adhere to the IETF MUD specification and 
are semantically correct are deployed to a network.

\subsection{Syntactic correctness}

A MUD profile consists of a YANG model which describes device-specific network behavior.
In the initial version of MUD, this model is serialized using JSON \cite{ietfMUD18}. 
JSON offers more compactness and readability, relative to XML. 
A MUD profile is limited to the serialization of only a few YANG modules (\eg  ietf-access-control-list) \cite{ietfMUD18}.
{\em MUDdy} will throw an invalid syntax exception when parsing a MUD profile if it encounters 
any schema beyond these permitted YANG modules.

In addition, {\em MUDdy} also rejects MUD profiles containing IP addresses (in particular those with local significance).
As per the IETF specification, publishers of MUD profiles are advised to
use the abstractions provided in the specification and avoid using IP addresses.
{\em MUDdy} will also reject a MUD profile if it contains actions other than "accept" or "drop".
\vspace{-3mm}

\subsection{Semantic correctness}

Checking a MUD policy's syntax partly verifies its correctness.
A syntactically correct policy must also be semantically correct for it to be well designed. 
For instance, it is necessary to check a MUD policy for
inconsistencies.

Inconsistencies in a MUD policy can stem from two sources; 
(a) due to overlapping rules with different access-control actions; and/or
(b)  due to overlapping rules with identical actions.
In the absence of rule ordering (as in the case of the IETF MUD specification), the former set of rules depict
ambiguous intent of the policy authors. We refer to such rules as {\em intent-ambiguous rules}.
Rule order is irrelevant for the latter type of rules which associate a clear (single) outcome.
We refer to such rules as {\em redundancies}.

Inconsistencies in a security policy can produce unintended consequences \cite{wool2004}.
So, correct MUD policy deployment demands the ability to check for inconsistencies accurately.
We do so by modeling MUD policies using metagraphs and then applying the readily available
metagraph algebras.

\subsubsection{Policy modeling}\label{sec:modeling}

\begin{figure}[t]
	\centering
	\includegraphics[scale=0.20]{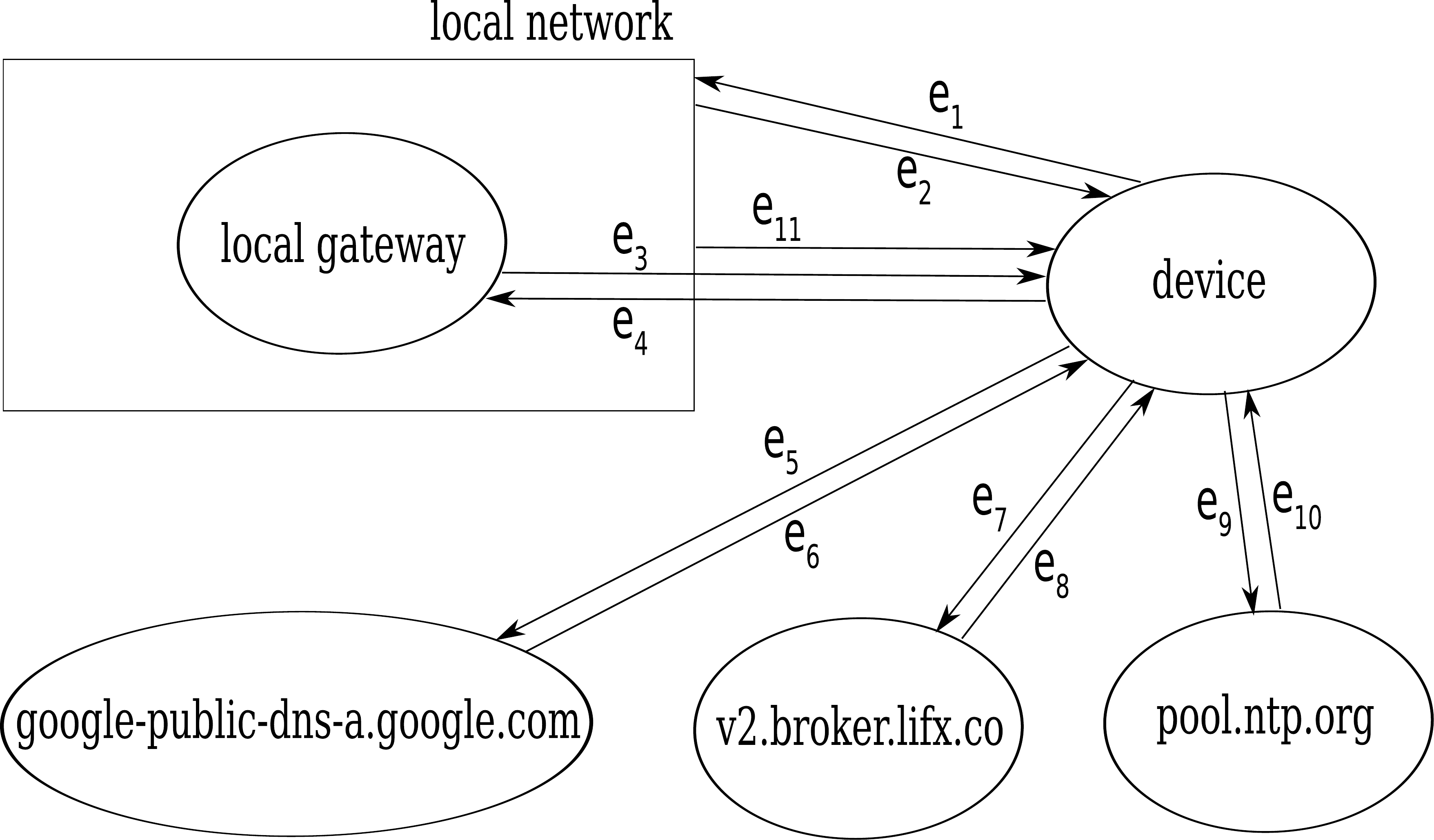}
	\caption{Metagraph model of a Lifx bulb's MUD policy. The policy describes permitted traffic flow behavior. 
		Each edge label depicts a set of propositions of the metagraph. For example $e_4$=$\{protocol=17,UDP.dport=53,UDP.sport=0-65535,action=accept \}$.}
	\label{fig:cmg_bulb}
\end{figure}

Access-control policies are commonly represented using the five-tuple: 
source/destination address, protocol, source/destination ports \cite{cisco2013,juniper2016,palo2017}.
We construct metagraph models for the MUD policies leveraging this idea.
A representative example from our case study is shown in \autoref{fig:cmg_bulb}.
Here, the source/destination addresses are represented by the labels {\em device}, {\em local-network}, {\em local-gateway} and a domain-name (\eg {\em pool.ntp.org}).
Protocol, ports and time are propositions of the conditional metagraph.

\subsubsection{Policy definition and verification}
\label{sec:definition}

We wrote {\em MGtoolkit} \cite{ranathunga2017} -- a package for implementing metagraphs -- 
to define our policy models. {\em MGtoolkit} is implemented in Python 2.7. 
The API allows users to instantiate metagraphs, apply metagraph operations and evaluate results.

{\em Mgtoolkit} provides a \texttt{ConditionalMetagraph} class which extends a \texttt{Metagraph} and 
supports proposition attributes in addition to variables.
A \texttt{ConditionalMetagraph} inherits the base properties and methods of a \texttt{Metagraph}
and additionally supports methods to check reachability properties and consistency properties.
We use the \texttt{ConditionalMetagraph} class to instantiate the MUD policy models in \autoref{sec:modeling}.
We then invoke the API methods to check policy properties such as consistency. 

Our verification of metagraph consistency uses {\em dominance} \cite{basu2007} which
can be introduced constructively as follows: 

\begin{defn}[Edge-dominant Metapath]
Given a metagraph $S$=$ \langle X, E \rangle$ for any two sets of elements
$B$ and $C$ in $X$, a metapath $M(B,C)$ is said to be edge-dominant if no proper
subset of $M(B,C)$ is also a metapath from $B$ to $C$.
\label{def:metapath1}
\end{defn}

\newlength{\figurewidthA}
\setlength{\figurewidthA}{0.31\textwidth}
\begin{figure*}[t!]
  \captionsetup{aboveskip=8pt}
  \centering
  \subfigure[Four rules indicated by (overlapping) rectangles.] 
  {
    \includegraphics[width=\figurewidthA]{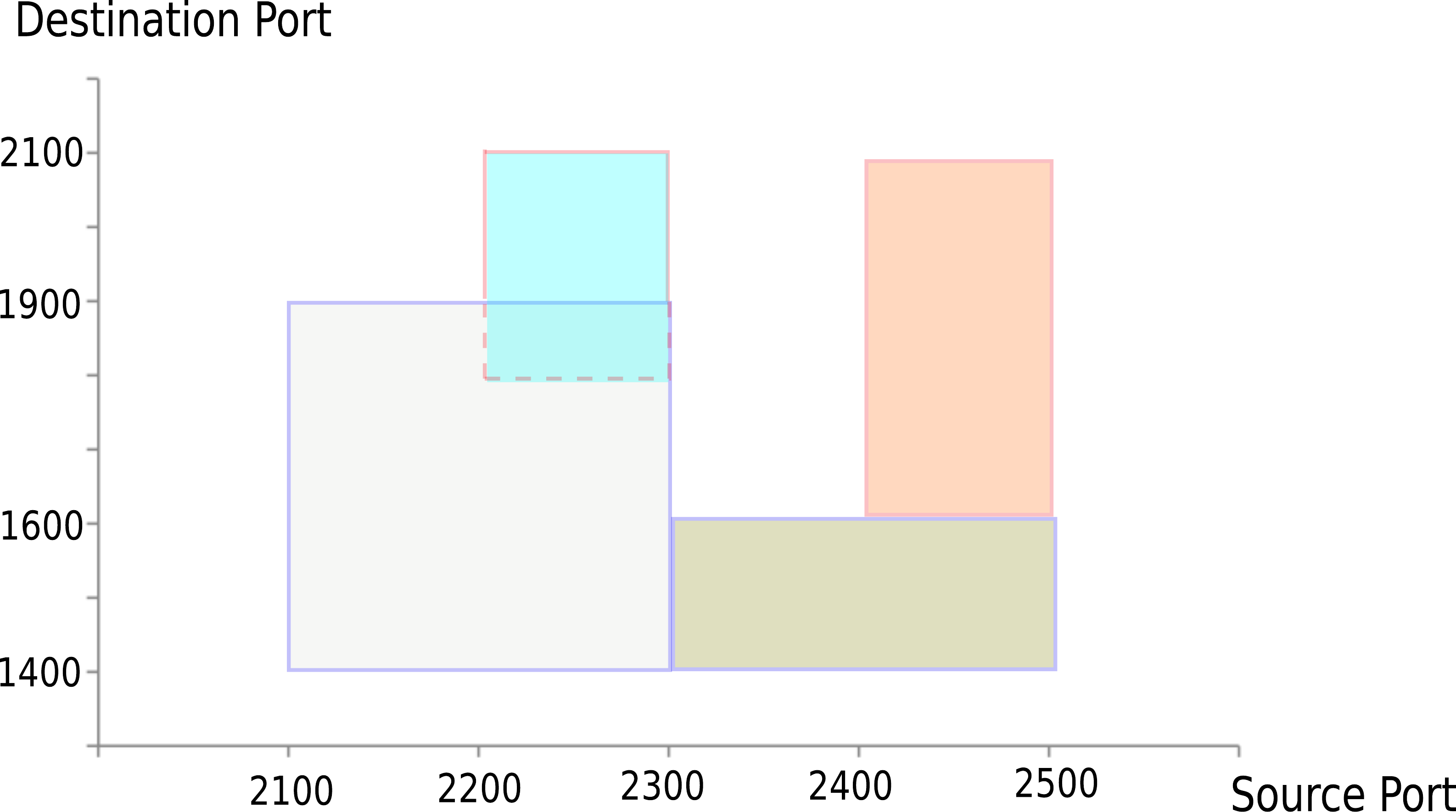}
    \label{fig:polygon1}
  }
  \subfigure[Five rules producing an equivalent policy to (a).] 
  {
    \includegraphics[width=\figurewidthA]{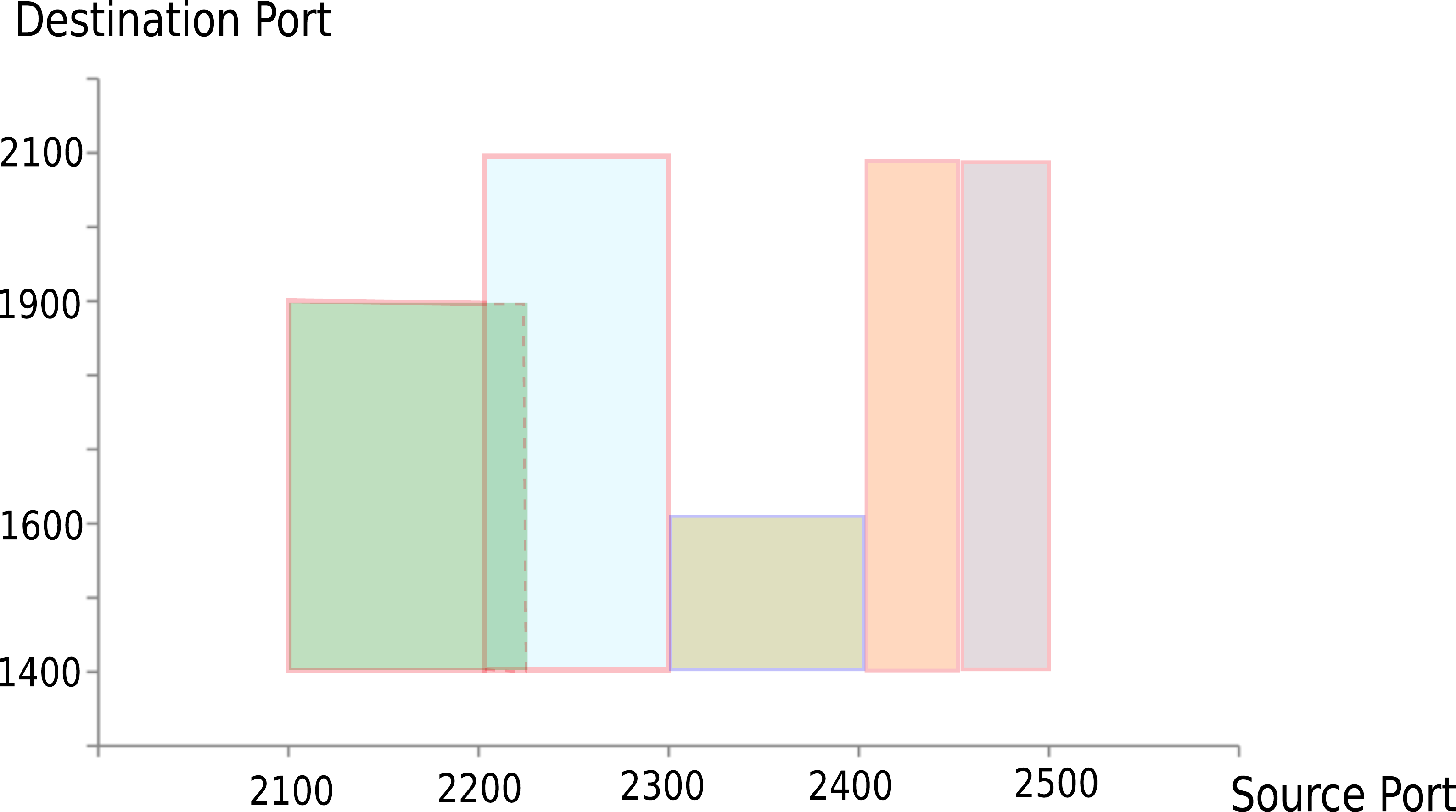}
    \label{fig:polygon2}
  }
  \subfigure[Horizontal partitions of polygon in (a) or (b).] 
  {
    \includegraphics[width=\figurewidthA]{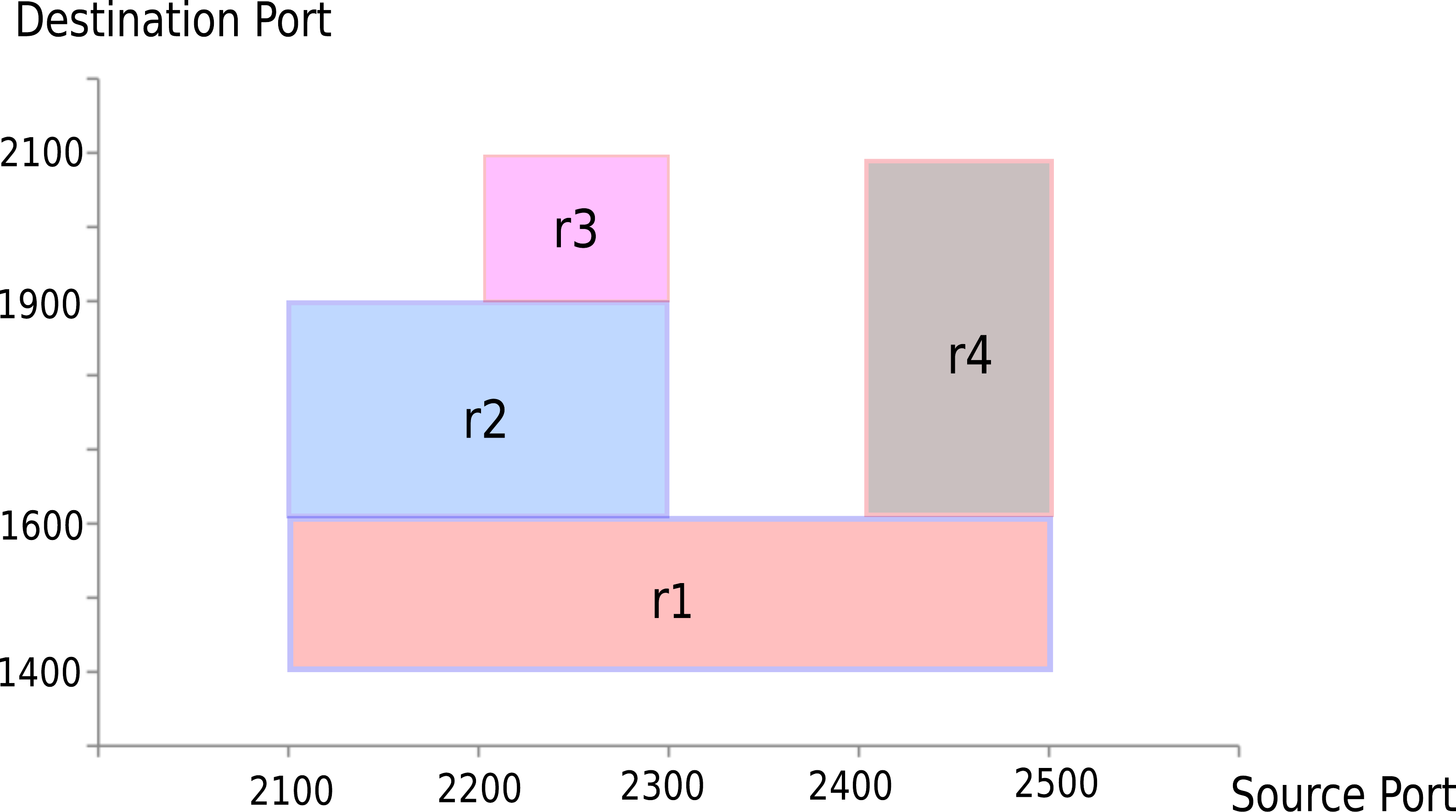}
    \label{fig:polygon3}
  }
  \caption{Canonicalisation of distinct rule sets of the same MUD policy. Rectangles indicate the packets allowed by a particular rule.}
  \label{fig:policy1}
 \vspace{-2mm}
\end{figure*}

\begin{defn}[Input-dominant Metapath]
Given a metagraph $S$=$ \langle X, E \rangle$ for
any two sets of elements $B$ and $C$ in $X$, a metapath $M(B,C)$ is said to be
input-dominant if there is no metapath $M'(B',C)$ such that $B' \subset B$.
\label{def:metapath2}
\end{defn}

In other words, edge-dominance (input-dominance) ensures that none of the
edges (elements) in the metapath is superfluous. Based on these
concepts, a dominant metapath can be defined as follows.
A non-dominant metapath indicates redundancy in the policy represented by the metagraph.

\begin{defn}[Dominant Metapath]
Given a metagraph $S$=$ \langle X, E \rangle$ for any two sets of elements
$B$ and $C$ in $X$, a metapath $M(B,C)$ is said to be dominant if it is both edge dominant
and input-dominant.
\label{def:metapath3}
\end{defn}

The potential `conflict set' of propositions in a metapath $M(B,C)$ can also be defined as follows:

\begin{defn}[Conflict-set of Propositions]
Given a conditional metagraph $S$=$ \langle X_v \cup X_p, E \rangle$ for any two sets of elements
$B$ and $C$ in $X$, a metapath $M(B,C)$ has the potential conflict set of propositions
given by $(\bigcup_{e \in M(B,C)} V_e ) \cap X_p$.
\label{def:metapath4}
\end{defn}

Once a conflict set is identified we can apply domain-specific knowledge
to determine intent-ambiguous rules in a MUD policy.
For instance, with access-control policies such rules occur when flows overlap and associate different response actions (\eg {\em accept} and {\em drop}).

\subsubsection{Compatibility with best practices}
\label{sec:bp_check}

Policy consistency checks partly verify if a MUD policy is semantically correct.
It may also be necessary to check MUD policy semantics against industry recommended
practices: \eg ANSI/ISA- 62443-1-1, for compliance. 
Doing so, is critical when installing an IoT device in a SCADA network.
where more restrictive practices are required to
prevent serious injury of people, or even death!

SCADA best practices offer a wide spectrum of security policies,
representative of various organizations, to compare our MUD policies against.
For instance, they include policies for the 
highly protected SCADA zone (which for instance runs a power plant)
as well as the more moderately restrictive Corporate zone.

We have investigated the problem of policy comparison using formal semantics, in the SCADA domain for firewall 
access-control policies \cite{ranathunga2016P}.
We adapt the methods and algebras developed there, to also
check MUD policies against SCADA best practices. 
Key steps enabling these formal comparisons are summarized below.

Equivalent MUD policies can be specified in many ways. 
Figures 5(a) and 5(b) illustrate the idea using TCP port filtering of single packets. 
Each rectangle indicates the allowed packets of a single rule. Combined the rules cover the same set of accepted packets.

An efficient approach to comparing these policies accurately is by deriving a unique, {\em canonical} representation of each MUD policy.
We dissect the polygon formed in our example policy into
horizontal partitions (Figure 5(c)), using a Rectilinear-Polygon 
to Rectangle conversion algorithm \cite{gourley1983}. Each partition is
chosen to provably guarantee its uniqueness. Canonical MUD policy
elements are derived by translating each partition back to a
rule and ordering the resulting rule-set uniquely in increasing
IP protocol number and source and destination port numbers.
We find a unique partition quickly rather than a guaranteed
minimal partition. The result is a deterministic, ordered set of
non-overlapping rules.

Canonicalisation of policies can be represented through a mapping $c: \Phi
\rightarrow \Theta$, where $\Phi$ is the policy space and $\Theta$ is the canonical space of
policies, in which all equivalent policies of $\Phi$ map to a
singleton. For $p^X, p^Y \in \Phi$, we note the following (the proof 
follows the definition)
\begin{lem}
	Policies $p^X \equiv p^Y$  iff $c(p^X)=c(p^Y)$.
\end{lem}

So, a MUD policy compliance check can be performed  by comparing canonical policy components.  For instance

\smallskip
\smallskip
\indent \indent Is $c ( p^{device \to controller} ) = c ( p^{SCADA \to Corp} )$ ? 

\smallskip
Another useful notation, linked to the goal of policy comparison,
is that policy $P^A$ {\em includes} policy $P^B$. 
Particularly in SCADA networks, the notation helps evaluate whether a MUD policy is
compliant with industry-recommended practices in \cite{stouffer2008,byres2005}. These guidelines 
specify potentially dangerous services (\eg HTTP) that should be {\em prohibited} from traversing inbound and/or outbound from the
(protected) SCADA zone. A violation here, means installing the IoT device increases the vulnerability of a SCADA zone to cyber attacks.

We indicate that a policy {\em complies} with another if it is more restrictive and define the following
\begin{defn}[Inclusion]
	A policy $p^X$ is {\em included} in $p^Y$ on $A$ iff $p^X(s)
	\in \{p^Y(s), \phi\}$, \ie $X$ either has the same effect as $Y$ on
	$s$, or denies $s$, for all $s \in A$. We denote inclusion by
	$p^X \subset p^Y$.
	\label{def:includes}
\end{defn}

A device MUD policy ($MP$) can now be checked against a best practice policy ($RP$) for compliance using an inclusion check

\smallskip
\indent \indent \indent \indent \indent  Is $p^{MP} \subset p^{RP}$ ?

\smallskip
The above approach
can likewise be used to check MUD policies against an organisation's internal security policy.
By doing so, we can ensure that IoT devices are plug and play enabled, only
in the compatible zones of a company's network.

\section{Evaluation of Results}\label{sec:eval}

\begin{figure*}[t!]
\centering
\includegraphics[scale=0.29]{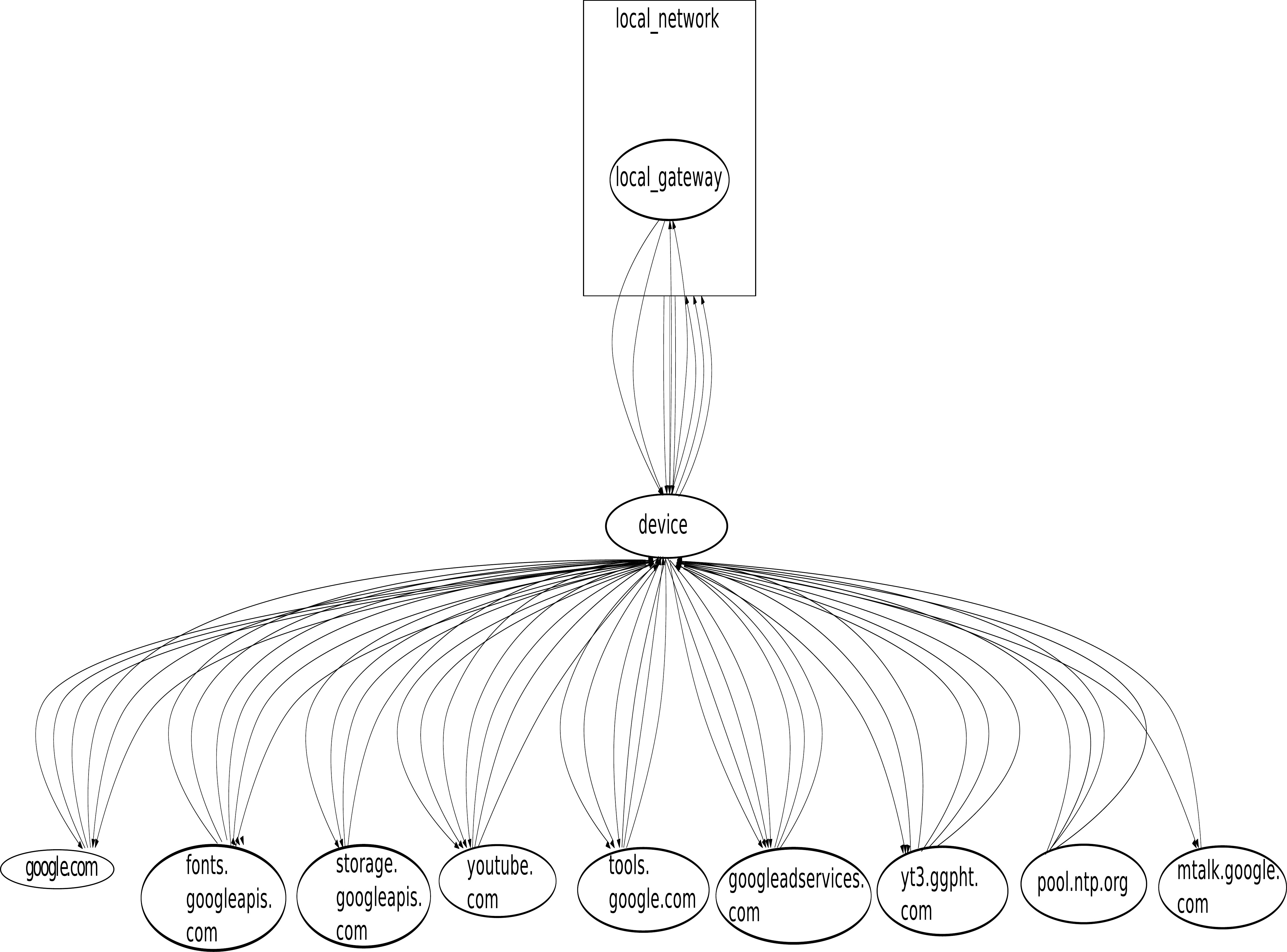}
\caption{Metagraph model of a Chrome cast device's MUD policy. The policy describes permitted traffic behavior.}
\label{fig:samsung}
\end{figure*}

\begin{table*}[t!]
\caption{MUD policy analysis summary for our test bed IoT devices ( {\em Safe to install?} indicates where in a network (\eg Enterprise Zone, SCADA Zone, DMZ) the device can be installed without violating best practices, LoC - Lines of Code, DMZ - Demilitarized Zone, Corp Zone - Enterprise Zone).} 
\centering 
\begin{adjustbox}{max width=0.9\textwidth}
\begin{tabular}{l r r c l r r} 
\hline\hline 
\multicolumn{1}{p{3.5cm}}{\centering Device name} & 
\multicolumn{1}{p{1cm}}{\centering \#MUD profile \\ rules} & 
\multicolumn{1}{p{1cm}}{\centering MUD profile \\ LoC} & 
\multicolumn{1}{p{2cm}}{\centering \#Redundant \\ rules } & 
\multicolumn{1}{p{2cm}}{\centering Safe to \\ install ?} &
\multicolumn{1}{p{1.5cm}}{\centering \% Rules violating \\ SCADA Zone} &
\multicolumn{1}{p{1.5cm}}{\centering \% Rules violating \\ Corp Zone}\\ [0.5ex] 
\hline 
Blipcare blood pressure monitor & 6 &182 & 0 & DMZ, Corp Zone & 50 & 0 \\ 
Netatmo weather station & 6 & 182 & 0 & DMZ, Corp Zone & 50 & 0\\ 
SmartThings hub & 10 & 298 & 0 & DMZ, Corp Zone & 60 & 0\\ 
Hello barbie doll & 12 & 368 & 0 & DMZ, Corp Zone & 33 & 0 \\ 
Withings scale & 15 & 430 & 4 & DMZ, Corp Zone & 33 & 0\\ 
Lifx bulb & 15 & 374  & 0 & DMZ, Corp Zone & 60 & 0 \\ 
Ring door bell & 16 & 522  & 0 & DMZ, Corp Zone & 38 & 0\\ 
Awair air quality monitor & 16 & 352 & 0 & DMZ, Corp Zone & 50 & 0 \\ 
Withings baby monitor & 18 & 430 & 0 & DMZ, Corp Zone & 28 & 0\\
iHome power plug & 17 & 368 & 0 & DMZ & 41 & 6\\ 
TPlink camera & 22 & 528 & 0 & DMZ & 50 & 4\\ 
TPlink plug & 25 & 766 & 0 & DMZ & 24 & 4\\ 
Canary camera & 26 & 786 & 0 & DMZ & 27 & 4 \\ 
Withings sleep sensor & 28 & 730 & 0 & DMZ & 29 & 4\\ 
Drop camera & 28 & 840 & 0 & DMZ & 43 & 11 \\ 
Net smoke sensor & 32 & 824 & 0 & DMZ & 25 & 3\\
Hue bulb & 33 & 979 & 0 & DMZ & 27 & 3 \\ 
Wemo motion detector & 35 & 799 & 0 & DMZ & 54 & 8\\ 
Triby speaker & 38 & 1046 & 0 & DMZ & 29 & 3\\ 
Netatmo camera & 40 & 1094 & 1 & DMZ & 28 & 2\\ 
Belkin camera & 46 & 1203 & 3 & DMZ & 52 & 11 \\ 
Pixstar photo frame & 46 & 1210 & 0 & DMZ & 48 & 28\\ 
August door bell camera & 55 & 744 & 9 & DMZ & 42 & 13 \\ 
Samsung smart camera & 62 & 1487 & 0 & DMZ & 39 & 19\\ 
Amazon echo speaker & 66 & 1800 & 4 & DMZ & 29 & 2 \\ 
HP printer & 67 & 1196  & 10 & DMZ & 25 & 9 \\ 
Wemo switch & 98 & 2523 & 3 & DMZ & 24 & 6\\ 
Chrome cast & 150 & 3863 & 24 & DMZ & 11 & 2 \\ 
\hline
\end{tabular}
\end{adjustbox}
\label{table:suc-summary} 
\end{table*}

\begin{figure*}[t!]
\centering
\includegraphics[scale=0.32]{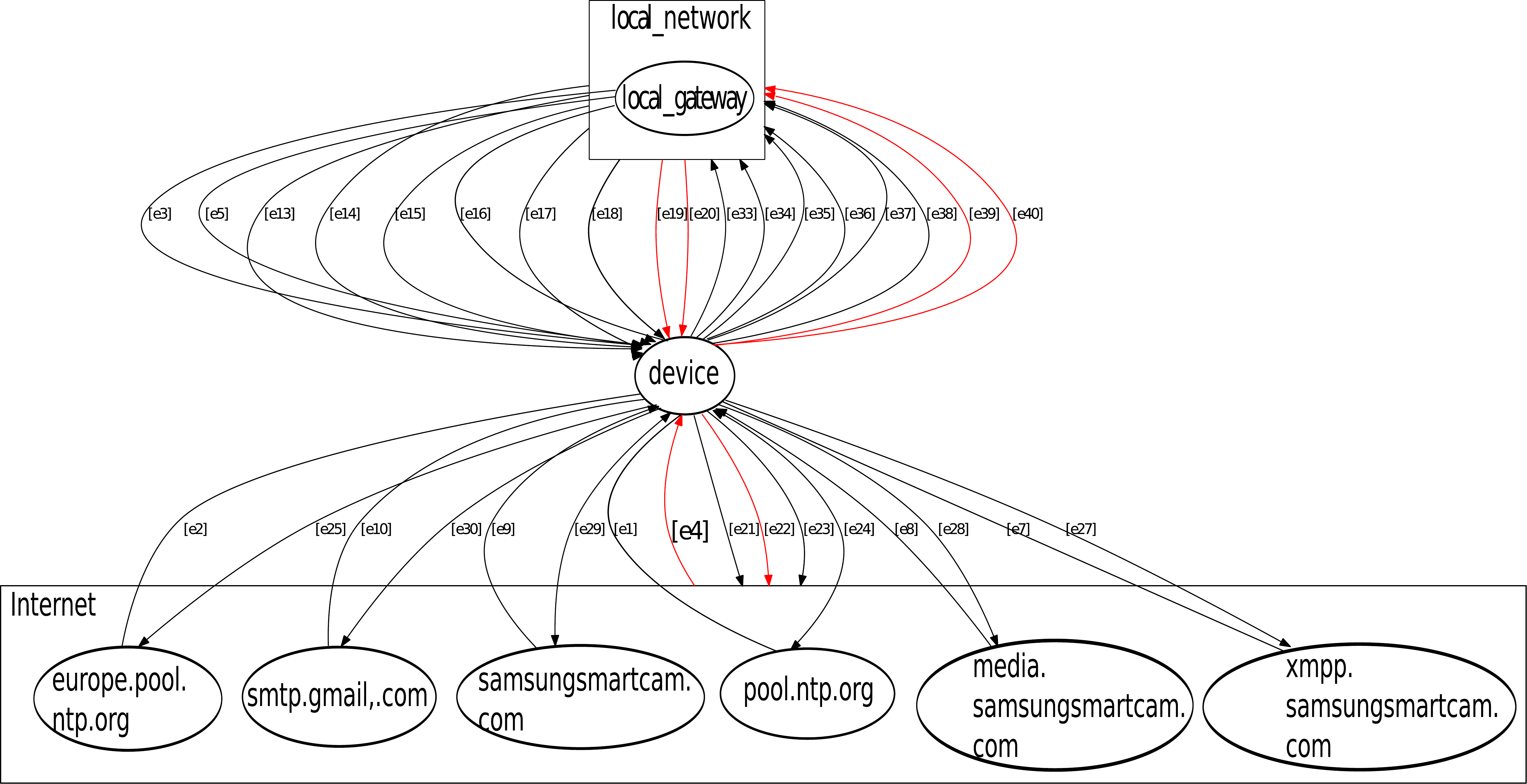}
\caption{Metagraph model of a Samsung smart camera's MUD policy. The policy describes permitted traffic (black edges) and denied traffic (red edges) behavior.
Each label edge depicts a set of propositions of the metagraph. For example $e_4$=$\{protocol=1, action=drop \}$.}
\label{fig:samsung}
\end{figure*}

We ran our system on a standard desktop computer (\eg Intel Core CPU 2.7-GHz computer with 8GB of RAM running
Mac OS X). It was used to generate the MUD files for 28 IoT devices in our test bed.
Each MUD file policy was then modeled using a conditional metagraph and instantiated using {\em MGtoolkit}.
A high-level summary of these MUD files and their metagraphs are given in \autoref{table:suc-summary}.

As the table shows, MUD file LoC is a good indicator of policy complexity.
For instance, a Chrome cast device has 3863 LoC in its MUD file while a Lifx bulb only has 374 LoC. 
The result indicates the former policy is more complex.
A metagraph helps to visualize this increase in policy complexity more easily.
The resemblance of a metagraph to a network allows one to exploit the pattern recognition capabilities of the
human visual cortex to better visualize policies.
For instance, compare the policy metagraph of a Chrome cast device in Figure 6 against that of the Lifx bulb in Figure 4.
One can easily understand the increase in size and complexity of the former policy relative to the latter, by simply inspecting the two metagraphs.

We identified inconsistencies in the MUD policies using the metagraph algebras defined in \autoref{sec:definition}.
There were no intent-ambiguous rules found since the MUD files were generated using an application whitelisting model (\ie restricting to accept rules).
However, redundancies are still possible and \autoref{table:suc-summary} shows the redundancies found.
For instance, there were three redundant rules in the Belkin camera's MUD policy (\autoref{table:suc-summary}).
These rules enabled ICMP traffic to the device from the local network as well as 
the local controller, making the policy inefficient.
We determined these redundancies by computing non-dominant metapaths in the policy metagraphs (as per Definition 4).

\autoref{table:suc-summary} also shows the results of our best practice 
compliance checks of the MUD policies. 
For instance, a Blipcare blood pressure monitor can be safely installed 
in the Demilitarized zone (DMZ) or the Corporate zone in a network, 
but not in the SCADA zone.
Installing the device in the SCADA zone would cause 
50\% of its MUD policy rules to violate the best practices (\autoref{table:suc-summary}),
exposing the zone to cyber attacks.
The violation is due to policy rules
which enable DNS from the device 
or enable the device to communicate with the Internet directly. 

In comparison, an Amazon echo speaker can only be safely installed in the DMZ of a network.
\autoref{table:suc-summary} shows that 29\% of the device's MUD policy rules violate the best practices if 
it's installed in the SCADA zone. Only 2\% of the rules violate if it's installed in the Corporate zone. 
The former violation stems from policy rules which for instance, enable
HTTP inbound to the device.
The latter stems from rules which enable ICMP inbound to the device from the Internet. 

{\em MUDdy}'s ability to pinpoint to MUD rules which cause best practice violations,
helps us to identify possible workarounds to achieve compliance.
For instance, with the Belkin camera,
local DNS servers and Web servers can be employed to
localize the device's DNS and Web communications
and overcome the violations.

\section{Discussion}\label{sec:discussion} 

Our use of an application whitelisting model in generating a MUD profile,
eliminates intent-ambiguous rules by design and reduces potential inconsistencies to
just redundancies. Thus, the effort required to check MUD policy consistency is reduced.
Moreover, the use of explicit `drop' rules requires considering rule order to
determine correct policy outcome.
The requirement increases policy specification complexity without any real benefit.

Drop rules also make policy visualization more difficult.
Consider the metagraph in Figure 7
which describes a MUD policy containing accept and drop rules.
Each metagraph edge describes an enabled traffic flow,
when a whitelisting model is adopted.
The absence of an edge implies two metagraph nodes don't communicate with one another.
But when drop rules are introduced, an edge now additionally describes prohibited traffic flows between the nodes.
This hinders easy visualization and understanding of a MUD policy.

Restricting to accept rules (with a default drop-all rule)
renders rule order irrelevant. 
A MUD policy holds the same semantics, irrespective of how rules are organized,
removing dependencies from the policy specification.
A policy author can then add or remove a rule without considering the complete rule set
and the potential interactions.
We recommend the IETF MUD specification be revised to only support explicit `accept' rules. 

Our suit of tools also allow one to check a MUD policy is compliant with an organizational policy, prior to deployment.
This ability reduces the effort required to acceptance-test an IoT device
because we need not test the device in network segments
where its MUD policy fails to comply with the organizational policy.

The IETF MUD specification also advises publishers of MUD profiles to use abstractions provided
and avoid using IP addresses. But upon clarification with the specification authors, 
we found that the use of public IP addresses is still permitted. This relaxation of the rule allows close
coupling of policy with network implementation, increasing its sensitivity to network changes.
A MUD policy describes IoT device behavior and should only change when 
its actual behavior alters and not due to changes to network implementation!
Hardcoded public IP addresses can also lead to DoS of target hosts.
A good example is the DoS of NTP servers at the University of Wisconsin due to hardcoded 
IP addresses in Netgear routers \cite{plonka2003}.
We recommend that support for explicit IP addresses (public or private)
be dropped from the MUD specification.

\section{Conclusion and Future Work}

The Internet of Things interconnects billions of every day devices to the Internet,
integrating the physical and cyber worlds.
But, increased interconnectivity also increases
the exposure of these devices to unwanted intrusions.
The IETF MUD proposal aims to reduce the threat surface
of an IoT device to what is intended by the manufacturer.

In this paper, we propose a suite of tools that allow to automatically
generate and formally verify IoT device MUD profiles,
to ensure the MUD policies are consistent and compatible with organizational policies.
We use these tools to demonstrate how MUD can reduce the effort needed
to secure IoT devices.

\addcontentsline{toc}{section}{List of Acronyms}
\begin{acronym}[TDMA]
        \setlength{\itemsep}{-\parsep}
        \acro{SCADA}{Supervisory Control and Data Acquisition}
        \acro{COTS}{Commodity Off-The-Shelf}
        \acro{TCP}{Transmission Control Protocol}
        \acro{IP}{Internet Protocol}
        \acro{ISA}{International Society for Automation}
        \acro{ASA}{Adaptive Security Appliance}
        \acro{PIX}{Private Internet eXchange}
        \acro{IOS}{IOS}
        \acro{WAN}{Wide Area Network}
        \acro{CLI}{Command Line Interface}
        \acro{DoS}{Denial of Service} 
        \acro{NMS}{Network Monitoring Station}
        \acro{ACLs}{Access Control Lists}
        \acro{ACE}{Access Control Entry}
        \acro{ACEs}{Access Control Entries}
        \acro{CZ}{Corporate Zone}
        \acro{SZ}{SCADA Zone}
        \acro{MZ}{Management Zone}
        \acro{FWZ}{Firewall Zone}
        \acro{IZ}{Internet Zone}  
        \acro{IDS}{Intrusion Detection Systems} 
        \acro{DMZ}{Demilitarised Zone}
        \acro{UDP}{User Datagram Protocol}
        \acro{DNS}{Domain Name System}
        \acro{HTTP}{Hypertext Transfer Protocol}
        \acro{HTTPS}{Hypertext Transfer Protocol Secure}
        \acro{FTP}{File Transfer Protocol}
        \acro{NTP}{Network Time Protocol}
        \acro{SSH}{Secure Shell Protocol}
        \acro{SMTP}{Simple Mail Transfer Protocol}
        \acro{SNMP}{Simple Network Management Protocol}
        \acro{ICMP}{Internet Control Message Protocol}
        \acro{DCOM}{Distributed Component Object Model}        
        \acro{NAT}{Network Address Translation}
        \acro{VPN}{Virtual Private Network}
        \acro{VPNs}{Virtual Private Networks}
        \acro{ANSI}{American National Standards Institute}
        \acro{OSI}{Open System Interconnection}
        \acro{SUC}{System Under Consideration}
        \acro{SUCs}{Systems Under Consideration}
        \acro{DPI}{Deep Packet Inspection}
        \acro{CN}{Carrier Network}
        \acro{IOS}{Internetwork Operating System}
        \acro{FWSM}{Firewall Services Module}
        \acro{PAT}{Port Address Translation}
        \acro{RBAC}{Role-Based Access Control}

\end{acronym}

\bibliographystyle{abbrv}
\bibliography{myBibliography} 

\begin{thebibliography}{10}

\bibitem{Insecam18}
{MUD maker}.
\newblock \url{http://www.insecam.org/en/bycountry/US/}, 2018.

\bibitem{basu2007}
A.~Basu and R.~Blanning.
\newblock {\em Metagraphs and their applications}, volume~15.
\newblock Springer Science \& Business Media, 2007.

\bibitem{f5Labs17}
S.~Boddy and J.~Shattuck.
\newblock {The Hunt for IoT: The Rise of Thingbots}.
\newblock Technical report, {F5 Labs}, July 2017.

\bibitem{byres2005}
E.~Byres, J.~Karsch, and J.~Carter.
\newblock {NISCC} good practice guide on firewall deployment for {SCADA} and
  process control networks.
\newblock {\em NISCC}, 2005.

\bibitem{cisco2013}
{Cisco Systems}.
\newblock {\em Cisco {ASA} Series {CLI} Configuration Guide, 9.0}.
\newblock Cisco Systems, Inc., 2013.

\bibitem{FCC16}
FCC.
\newblock {Federal Communications Comssion Response 12-05-2016}.
\newblock \url{https://goo.gl/JdLofa}, 2016.

\bibitem{gourley1983}
K.~D. Gourley and D.~M. Green.
\newblock Polygon-to-rectangle conversion algorithm.
\newblock {\em IEEE CGA}, pages 31--32, 1983.

\bibitem{mudgenerator}
A.~Hamza.
\newblock {MUDgee}.
\newblock \url{https://github.com/ayyoob/mudgee}, 2018.

\bibitem{Dyn16}
S.~Hilton.
\newblock {Dyn Analysis Summary Of Friday October 21 Attack}.
\newblock \url{https://goo.gl/mCdQUF}, 2016.

\bibitem{juniper2016}
{Juniper Networks, Inc.}
\newblock {\em Getting Started Guide for the Branch {SRX} Series}.
\newblock 1133 Innovation Way, Sunnyvale, CA 94089, USA, 2016.

\bibitem{ietfMUD18}
E.~Lear, R.~Droms, and D.~Romascanu.
\newblock Manufacturer usage description specification (work in progress).
\newblock Internet-Draft draft-ietf-opsawg-mud-18, IETF Secretariat, January
  2018.

\bibitem{IoTSnp17}
F.~Loi, A.~Sivanathan, H.~H. Gharakheili, A.~Radford, and V.~Sivaraman.
\newblock Systematically evaluating security and privacy for consumer iot
  devices.
\newblock In {\em Proc. ACM IoT S\&P}, Dallas, Texas, USA, Nov 2017.

\bibitem{Shodan}
J.~Matherly.
\newblock {Shodan}.
\newblock [Online]. Available: \url{https://www.shodan.io/}, 2018.

\bibitem{Survey17}
D.~M. Mendez, I.~Papapanagiotou, and B.~Yang.
\newblock Internet of things: Survey on security and privacy.
\newblock {\em CoRR}, abs/1707.01879, 2017.

\bibitem{ENISA17}
E.~U. A.~F. Network and I.~Security.
\newblock {Communication network dependencies for ICS/SCADA Systems}.
\newblock
  \url{https://www.enisa.europa.eu/publications/ics-scada-dependencies}, 2017.

\bibitem{NIST16}
NIST.
\newblock {Systems Security Engineering}.
\newblock \url{https://goo.gl/Qo9GfD}, 2016.

\bibitem{DHS16}
U.~D. of~Homeland~Security.
\newblock {Strategic Principles For Securing the Internet of Things (IoT)}.
\newblock \url{https://goo.gl/PaXbc4}, 2016.

\bibitem{palo2017}
{Palo Alto Networks, Inc.}
\newblock {\em PAN-OS Administrator's Guide, 8.0}.
\newblock 4401 Great America Parkway, Santa Clara, CA 95054, USA, 2017.

\bibitem{plonka2003}
D.~Plonka.
\newblock {Flawed Routers Flood University of Wisconsin Internet Time Server}.
\newblock \url{www.pages.cs.wisc.edu/~plonka/netgear-sntp/}, 2013.

\bibitem{ranathunga2017}
D.~Ranathunga, H.~Nguyen, and M.~Roughan.
\newblock Mgtoolkit: A python package for implementing metagraphs.
\newblock {\em SoftwareX}, 6:91--93, 2017.

\bibitem{ranathunga2016P}
D.~Ranathunga, M.~Roughan, P.~Kernick, and N.~Falkner.
\newblock Malachite: Firewall policy comparison.
\newblock In {\em IEEE Symposium on Computers and Communication (ISCC)}, pages
  310--317, June 2016.

\bibitem{ranathunga2016G}
D.~Ranathunga, M.~Roughan, P.~Kernick, N.~Falkner, H.~Nguyen, M.~Mihailescu,
  and M.~McClintock.
\newblock Verifiable policy-defined networking for security management.
\newblock In {\em SECRYPT}, pages 344--351, 2016.

\bibitem{ranathunga2016T}
D.~Ranathunga, M.~Roughan, H.~Nguyen, P.~Kernick, and N.~Falkner.
\newblock Case studies of scada firewall configurations and the implications
  for best practices.
\newblock {\em IEEE Transactions on Network and Service Management},
  13:871--884, 2016.

\bibitem{schmidt2008}
M.~Schmidt.
\newblock The sankey diagram in energy and material flow management.
\newblock {\em Journal of industrial ecology}, pages 82--94, 2008.

\bibitem{SmartCity17}
A.~Sivanathan, D.~Sherratt, H.~H. Gharakheili, A.~Radford, C.~Wijenayake,
  A.~Vishwanath, and V.~Sivaraman.
\newblock Characterizing and classifying iot traffic in smart cities and
  campuses.
\newblock In {\em Proc. IEEE INFOCOM workshop on SmartCity}, Atlanta, Georgia,
  USA, May 2017.

\bibitem{sivaraman2016smart}
V.~Sivaraman, D.~Chan, D.~Earl, and R.~Boreli.
\newblock Smart-phones attacking smart-homes.
\newblock In {\em Proceedings of the 9th ACM Conference on Security \& Privacy
  in Wireless and Mobile Networks}, pages 195--200. ACM, 2016.

\bibitem{stouffer2008}
K.~Stouffer, J.~Falco, and K.~Scarfone.
\newblock Guide to {I}ndustrial {C}ontrol {S}ystems ({I}{C}{S}) security.
\newblock {\em NIST Special Publication}, 800(82):16--16, 2008.

\bibitem{CiscoReport18}
C.~Systems.
\newblock {Cisco 2018 Annual Cybersecurity Report}.
\newblock Technical report, 2018.

\bibitem{wool2004}
A.~Wool.
\newblock A quantitative study of firewall configuration errors.
\newblock {\em IEEE Computer}, 37(6):62--67, 2004.

\bibitem{wool2010}
A.~Wool.
\newblock Trends in firewall configuration errors: Measuring the holes in
  {S}wiss cheese.
\newblock {\em IEEE Internet Computing}, 14(4):58--65, 2010.

\bibitem{SonyCam}
P.~World.
\newblock {Backdoor accounts found in 80 Sony IP security camera models}.
\newblock \url{https://goo.gl/UUvc2x}, 2018.

\end{thebibliography}

\end{document}